\documentclass[english]{bullsrsl}

\usepackage[latin1]{inputenc}

\usepackage[T1]{fontenc}

\usepackage{natbib}          

\usepackage{graphicx}

\begin{document}
\title{Going from 3D common-envelope simulations to fast 1D simulations}





\author[affil={1,2}, corresponding]{Vincent~A.}{Bronner}
\author[affil={1,3}]{Fabian~R.~N.}{Schneider}
\author[affil=4]{Philipp}{Podsiadlowski}
\author[affil={1,5}]{Friedrich~K.}{R\"opke}

\affiliation[1]{Heidelberg Institute for Theoretical Studies, Heidelberg}
\affiliation[2]{Universit\"at Heidelberg, Department of Physics and Astronomy, Heidelberg}
\affiliation[3]{Zentrum f\"ur Astronomie der Universit\"at Heidelberg, Astronomisches Rechen-Institut, Heidelberg}
\affiliation[4]{University of Oxford, St Edmund Hall, Oxford}
\affiliation[5]{Zentrum f\"ur Astronomie der Universit\"at Heidelberg, Institut f\"ur Theoretische Astrophysik, Heidelberg}


\correspondance{vincent.bronner@h-its.org}


\maketitle

\begin{abstract}
One-dimensional (1D) methods for simulating the common-envelope (CE) phase offer advantages over three-dimensional (3D) simulations regarding their computational speed and feasibility. We present the 1D CE method from \citet{Bronner2024}, including the results of the CE simulations of an asymptotic giant branch star donor. We further test this method in the massive star regime by computing the CE event of a red supergiant with a neutron-star mass and a black-hole mass companion. The 1D model can reproduce the orbital evolution and the envelope ejection from 3D simulations when choosing suitable values for the free parameters in the model. The best-fitting values differ from the expectations based on the low mass simulations, indicating that the free parameters depend on the structure of the giant star. The released recombination energy from hydrogen and helium helps to expand the envelope, similar to the low-mass CE simulations.
\end{abstract}


\section{Introduction}
The common-envelope (CE) evolution is one of the most challenging phases in the theory of binary star evolution. A range of both time and length scales involved in the CE phase make it difficult for any numerical simulations \citep{Roepke2023}. Yet, a good understanding of the CE phase is important for predicting the post-CE evolution. This is true, especially for the evolution of massive binary stars going through the CE phase, as this is one possible formation channel of gravitational-wave mergers and potentially the dominant channel for binary neutron-star mergers \citep[e.g.][]{Tutukov1993, Belczynski2002, Tauris2017, VignaGomez2018, Mandel2022}.

To date, three-dimensional (3D) hydrodynamic simulations are a key tool for studying CE events \citep{Ricker2008, Passy2012, Ohlmann2016a, Lau2022a}. However, these simulations are computationally expensive and run easily for $\sim 10^5$ core-hours. To overcome these large computational costs, one-dimensional (1D) models for CE simulation have been proposed \citep[see][]{Meyer1979, Podsiadlowski2001, Clayton2017, Fragos2019, Hirai2022}, either to speed up the simulations, or because the computational power was not available at that time, or to cover timescales that are not feasibly by the 3D hydrodynamical calculations.

\citet{Bronner2024} proposed a 1D CE method that reduces the computational cost to less than 10 core-hours. It has been shown that this method works well for CE events of a $0.97\,\mathrm{M}_\odot$ asymptotic giant branch (AGB) star with companions with mass ratios of 0.25, 0.5, and 0.75. Here, we demonstrate that this method is also applicable for CE events that include massive stars, such as red supergiants (RSG) as presented in \citet{Moreno2022} and revised in \citet{Vetter2024}. First, we summarize the methods and prerequisites of the 1D CE simulations in Section~\ref{sec:meth}. Then we present the results in Section~\ref{sec:res}, before giving a summary and a short outlook into future plans in Section~\ref{sec:conc}.

\section{Methods}\label{sec:meth}
The methods used for this work are presented in detail in \citet{Bronner2024}. Nonetheless, we will give a short overview of the 1D CE method in Section~\ref{sec:meth_summary}, before introducing the initial model used for the CE simulations of the RSG donor in Section~\ref{sec:meth_initial_model}

\subsection{1D CE method summary}\label{sec:meth_summary}
\begin{figure}
    \centering
    \includegraphics[width=8cm]{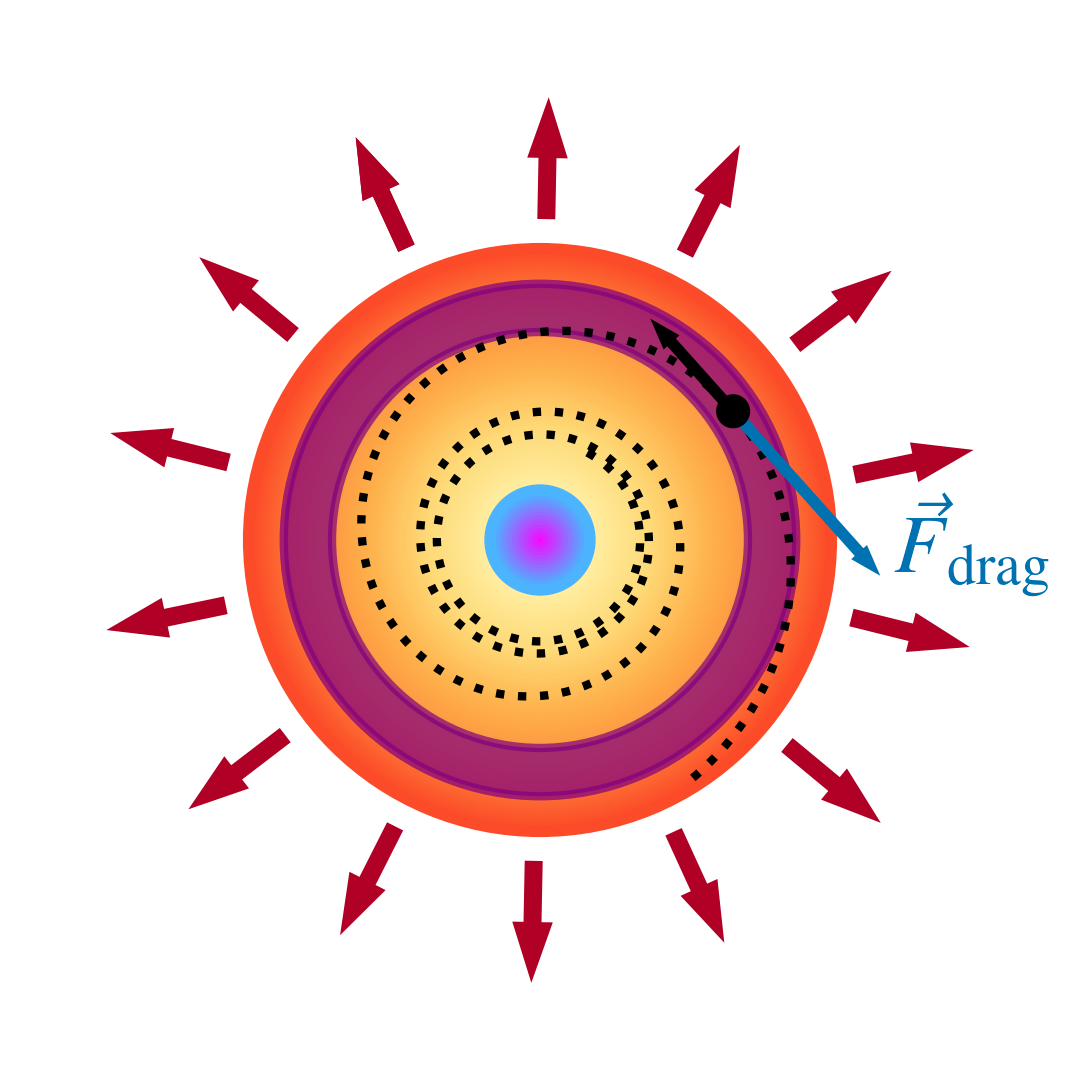}
    \bigskip
    \begin{minipage}{12cm}
    \centering
        \caption{Cartoon illustrating the 1D CE method. The companion (black dot) is spiraling inside the giant star. The black and blue arrows indicate the relative velocity of the companion and the drag force respectively. The heating zone is highlighted by the purple ring.}
        \label{fig:catoon}
    \end{minipage}
\end{figure}

The CE phase is modeled using the hydrodynamic capabilities of the 1D stellar-evolution code \texttt{MESA} \citep{Paxton2011, Paxton2013, Paxton2015, Paxton2018, Paxton2019}. We assume that the CE is spherically symmetric and centered on the core of the giant star. In parallel to the \texttt{MESA} computation, we integrate the equations of motion of the binary star to obtain the orbit of the giant star and the companion (Fig.~\ref{fig:catoon}). For the companion star, we add a gravitational drag term to the equations of motions \citep{Kim2007, Kim2010}. Initially, the companion is placed in a circular orbit just below the surface of the giant star. This setup does not account for any pre-CE mass-transfer phases that might have altered the structure of the giant star \citep{Blagorodnova2021, Renzo2023}. Because of the drag force, orbital energy is released from the binary star system. The released orbital energy is added to the envelope of the giant star by heating the layers around the companion (later called the heating zone). The radial extent of the heating zone is estimated using the Bondi-Lyttleton-Hyole accretion model \citep{Hoyle1941, Bondi1944}. As the envelope is expanding during the CE phase, we remove any surface layers of the CE, that are unbound (i.e.,~their expansion velocity is larger than the escape velocity).

Two free parameters are introduced to the model. The drag-force parameter $C_\mathrm{d}$ scales the strength of the drag force on the companion, that is $F_\mathrm{d} = C_\mathrm{d} \times ...$, where the entire expression for the drag force $F_\mathrm{d}$ is given in detail in \citet{Bronner2024}. The heating parameter $C_\mathrm{h}$ varies the radial extent of the heating zone by modifying the accretion radius $R_\mathrm{a}$ via $R_\mathrm{a} = C_\mathrm{h} \times ...$, again referring to \citet{Bronner2024} for the details in the heating prescription and the determination of the radial extend of the heating zone. We vary both parameters to fit the evolution of the orbital separation (spiral-in curve) and the fraction of ejected envelope mass from our 1D simulations to the results from 3D hydrodynamical simulations. The 3D simulations are started at 60\% of the orbital separation at which the giant star would fill its Roche lobe. In particular, this means that the companion is still outside the envelope of the giant star. The 1D simulations can only be started once the companion is inside the envelope, which is responsible for the drag force and the release of orbital energy. The difference in initial separation is compensated by offsetting the time between the 1D and 3D simulation when comparing the spiral-in curve and the envelope ejection.

\subsection{Initial model for RSG CE simulations}\label{sec:meth_initial_model}
We aim to fit our 1D CE simulations to the results of 3D CE simulations of an RSG with either a neutron-star (NS) mass or a black-hole (BH) mass companion presented in \citet{Vetter2024}. First, we reconstruct the initial 1D stellar model used to start the 3D simulations. We evolve a $10\,\mathrm{M}_\odot$ star at solar metallicity from the zero-age main sequence to the RSG phase and stop the simulations once the star reaches a radius of $\sim 390\, \mathrm{R}_\odot$, to match the radius of the progenitor in \citet{Vetter2024}. The mass of the final RSG model is $9.25\,\mathrm{M}_\odot$ compared to $9.4\,\mathrm{M}_\odot$ in \citet{Vetter2024}. This deviation arises because slightly different wind mass-loss prescriptions were used. However, the deviations in the mass distribution are less than $2 \%$ throughout the entire star, ensuring an alike starting condition for the CE simulations.

\section{Results}\label{sec:res}
In this section, we present the results of the CE simulations using the 1D method. First, we will summarize the main results of the simulations with the AGB donor in Section~\ref{sec:res_agb}. Then, we present the results of the CE simulations with the RSG donor in Section~\ref{sec:res_rsg}. Finally, we compare the results of the AGB and RSG simulations in Section~\ref{sec:res_comp}.

\subsection{AGB-star donor}\label{sec:res_agb}
The simulations with the AGB stars donor are analyzed in detail in \citet{Bronner2024}. We will only give a summary of the main findings in this section.

We compute the CE evolution of an AGB star with a companion with mass ratios $q = 0.25$, 0.5, and 0.75 and compare the results to the 3D simulations by \citet{Sand2020}. We find well-fitting simulations for $q=0.25$ ($C_\mathrm{d} = 0.23,\,C_\mathrm{h}=4.00$) and $q = 0.5$ ($C_\mathrm{d} = 0.30,\,C_\mathrm{h}=1.30$). For $q=0.75$, we are unable to find values for $C_\mathrm{d}$ and $C_\mathrm{h}$ such that the 1D simulation is able to reproduce the 3D simulation. Regardless of the mass ratio, we need two free parameters in our model to reproduce both the spiral-in behavior and the envelope ejection process from the 3D simulations. Constraining our model to just one parameter is not possible. We find that not just the strength of the drag force but also where the energy is added makes a difference in the orbital separation and most notably in the timing and the amount of envelope ejection. Since we use the models for the drag force \citep[see][]{Kim2007,Kim2010} and the heating \citep[see][]{Hoyle1941, Bondi1944} beyond the intended use, we need to recalibrate both using the free parameters. First, the companion is moving through an envelope with a density gradient, and the companion is not on a circular orbit \citep[see][]{Kim2007, Kim2010}. Additionally, the released orbital energy in the 3D simulations is not directly converted to thermal energy but rather to kinetic energy and subsequently to thermal energy.  

When analyzing the envelope structure and the expansion and ejection process, we find that the released recombination energy of hydrogen and helium aids the envelope expansion. The recombination energy of hydrogen plays a significant role in the expansion.

\subsection{RSG-star donor}\label{sec:res_rsg}

\begin{figure}
    \centering
    \includegraphics[width=12cm]{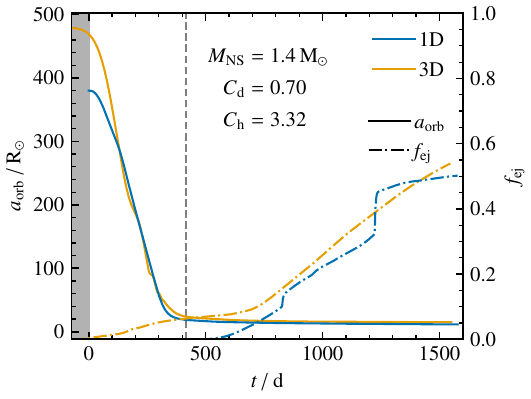}
    \bigskip
    \begin{minipage}{12cm}
    \centering
        \caption{Best-fitting simulation of the RSG with the NS-mass companion using $C_\mathrm{d} = 0.70$ and $C_\mathrm{h} = 3.32$. Full lines show the orbital separation $a_\mathrm{orb}$ and the dash-dotted lines show the fraction of ejected envelope mass $f_\mathrm{ej}$. The orange lines correspond to the results from \citet{Vetter2024}. The grey-shaded region indicates the time offset of $\Delta t = -73.6\,\mathrm{d}$ used to account for the different starting conditions between the 1D and 3D simulations. The end of the plunge-in phase is marked by the vertical dashed line.}
        \label{fig:NS}
    \end{minipage}
\end{figure}

We simulate the CE event of a $9.25\,\mathrm{M}_\odot$ RSG with a NS mass companion of $M_\mathrm{NS} = 1.4\, \mathrm{M}_\odot$. The comparison to the results of 3D simulations by \citet{Vetter2024} yield the best-fitting parameters $C_\mathrm{d} = 0.70$ and $C_\mathrm{h}=3.32$ (Fig.~\ref{fig:NS}). We find that the spiral-in curve of the 1D simulation is in excellent agreement with the 3D simulation, both during the plunge-in phase (i.e., the phase during which the orbital separation changes dynamically) and the post plunge-in phase. The rate of change of the orbital separation in the 1D simulation matches that of the 3D simulation, even though they are started at different initial orbital separations. As a result, the orbital energy of the binary star is released at a similar rate in the 1D and 3D simulations. 

The dynamical plunge-in phase terminates after a few orbits at around $t=400\, \mathrm{d}$. We find similar post-plunge-in separations in both the 1D and the 3D simulations. At $t=1500\, \mathrm{d}$, the orbital separation is $12.8\, \mathrm{R}_\odot$ in the 1D simulation compared to $15.2 \,\mathrm{R}_\odot$ in the 3D simulation. The 1D simulation is stopped at $t=1577 \,\mathrm{d}$ because of unrecoverable numerical difficulties. They originate in the heating zone, which has been cleared of matter resulting in a low-density environment.

When looking at the fraction of ejected envelope mass, we find that in the 1D simulation, the envelope ejection only starts at around $t = 500 \,\mathrm{d}$. By the time of $t=850\,\mathrm{d}$, the 1D simulation catches up with the 3D simulation, and the envelope is ejected at a similar rate. The reason for this might be that the orbital energy is released at a similar rate in both the 1D and 3D simulations. However, it was shown in \citet{Bronner2024} that the amount of ejected envelope is direly influenced by the heating parameter $C_\mathrm{h}$, which only changes the radial extent of the heated layer but not the energy deposition rate. There are two ejection busts during the CE phase. They originate from formally unbound layers just below the surface of the CE. Matter piles up in these layers and once a critical mass is accumulated, they reach the surface and get ejected instantaneously. We speculate, that this is an artifact of the surface boundary condition used. At around $t=1300 \, \mathrm{d}$, the envelope ejection rate is much lower than in the 3D simulation. It is not clear if that trend will continue or if another period of larger mass ejection rate will follow, as we are not able to run the simulation for longer.

\begin{figure}
    \centering
    \includegraphics[width=12cm]{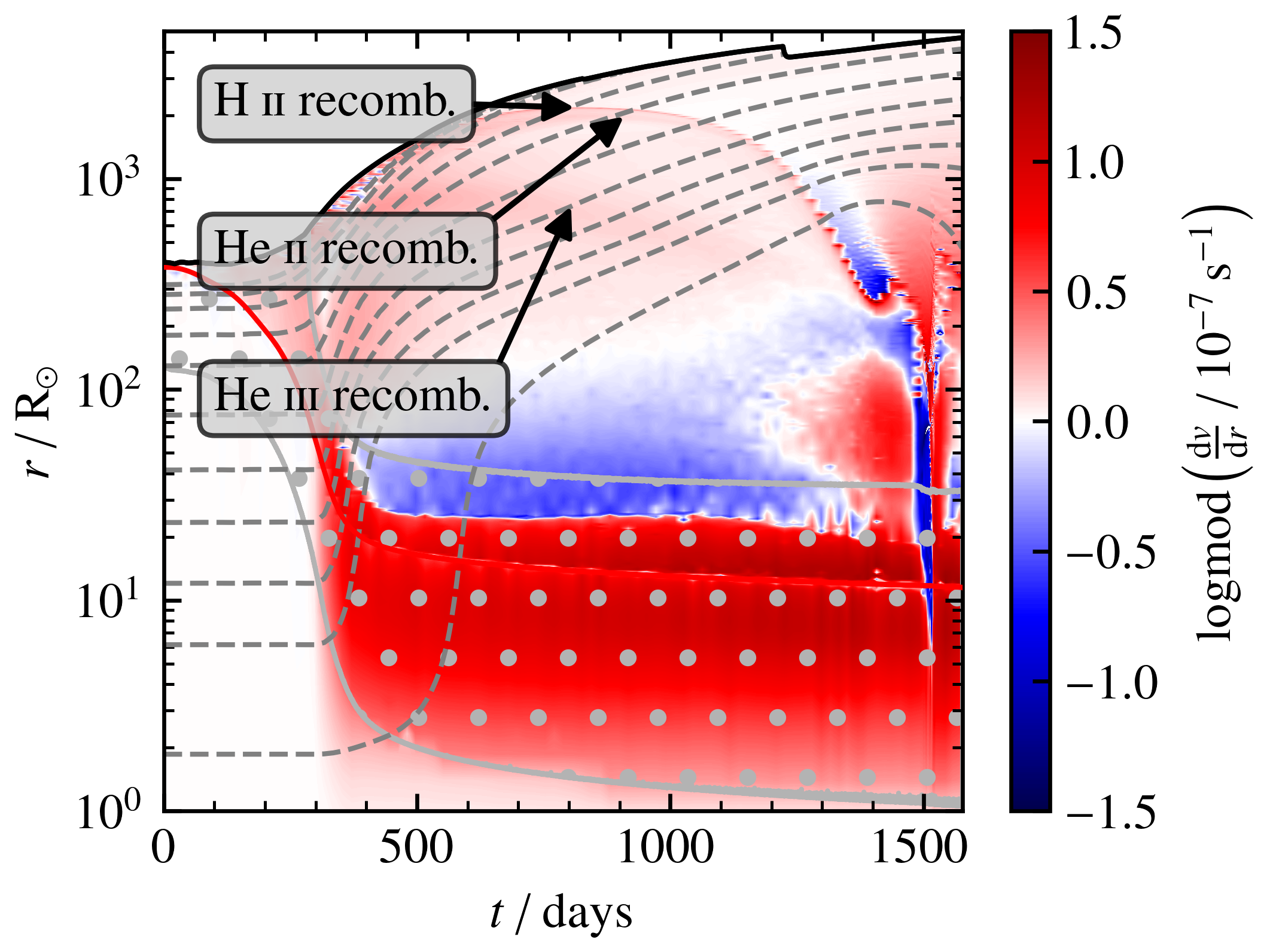}
    \bigskip
    \begin{minipage}{12cm}
    \centering
        \caption{Structure evolution of the RSG envelope with the NS mass companion showing the velocity divergence $\mathrm{d}v/\mathrm{d}r$. The red sold line shows the location of the companion inside the envelope. The dotted region corresponds to the heating zone. The dashed lines show lines of constant envelope mass fraction of 0.95, 0.9, 0.8, 0.6, 0.4, 0.2, 0.1, 0.06, 0.04, 0.03, and 0.02 from the surface towards the center. For better visualization, the logmod transformation, $\mathrm{logmod}(x) = \mathrm{sign}(x) \log_{10}(|x|+1)$, is used to show the velocity divergence \citep{John1980}.}
        \label{fig:kipp_dv_dr}
    \end{minipage}
\end{figure}

The envelope structure of the RSG with the NS mass companion is shown in Fig.~\ref{fig:kipp_dv_dr}. The velocity divergence $\mathrm{d}v/\mathrm{d}r$ is a measure of whether the envelope is expanding in an accelerated way (red color in Fig.~\ref{fig:kipp_dv_dr}) or in a decelerated way (blue color in Fig.~\ref{fig:kipp_dv_dr}). We find positive velocity divergence in most parts of the heated zone around the NS mass companion, with the largest values reached during the post-plunge-in phase. This is expected because we increase the internal energy in these layers, which results in the expansion of the envelope. During the post-plunge-in phase, the density in the heating zone is much lower than during the plunge-in phase, causing a larger specific energy deposition rate. Layers of positive velocity divergence in the outer layers of the envelope can be associated with the recombination layers of hydrogen and helium. Hydrogen and singly-ionized helium (He~{\sc ii}) recombine at similar radii, but vastly different optical depths ($\log_{10}\tau \sim 1-4$ for hydrogen and $\log_{10}\tau \sim 4.5$ for He~{\sc ii}). Double-ionized helium (He~{\sc iii}) recombines further inside the envelope at an optical depth of $\log_{10}\tau \sim 5$.  All three recombination layers contribute towards the envelope expansion as indicated by the positive velocity divergence. Additionally, we find a steepening of the slope of lines of constant envelope mass fraction in the He~{\sc iii} recombination layer.

\begin{figure}
    \centering
    \includegraphics[width=12cm]{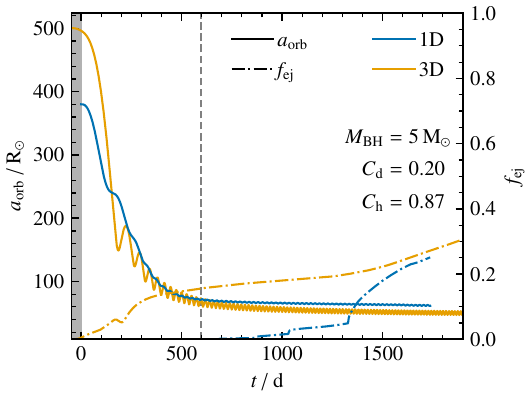}
    \bigskip
    \begin{minipage}{12cm}
    \centering
        \caption{Same as Fig.~\ref{fig:NS} but with the BH mass companion, but this time using $C_\mathrm{d} = 0.20$ and $C_\mathrm{h} = 0.87$. We apply a time offset of $\Delta t = -47.8 \, \mathrm{d}$.}
        \label{fig:BH}
    \end{minipage}
\end{figure}

We repeat the simulation with a BH mass companion of $M_\mathrm{BH} = 5 \, \mathrm{M}_\odot$ and find best-fitting parameters of $C_\mathrm{d}= 0.20$ and $C_\mathrm{h} = 0.87$ (Fig.~\ref{fig:BH}). The orbital separation during the plunge-in phase is well reproduced, although the eccentricity in the 1D simulation is lower compared to the 3D simulation. It has been shown that dynamical friction can increase the eccentricity of the orbit, depending on the density structure \citep[e.g.][]{Szoelgyen2022}. We initiate the simulation on a circular orbit (i.e.~zero eccentricity). Because the energy deposition timescale by the drag force is much smaller than the orbital period, a change in eccentricity is expected. The lower initial separation leads to a shallower plunge-in during the first orbit and subsequently results in a more steady orbital decline rate. Nonetheless, the rate of change of the orbital separation of the 1D and the 3D simulations are comparable when averaging over one orbit. 

During the post-plunge-in phase, we find a larger orbital separation in the 1D simulation compared to the 3D simulation. In addition, the orbital separation in the 1D simulation declines at a lower rate, and the eccentricity is smaller. Possible reasons for this are a different density structure of the CE around the companion at this stage, that changes the strength of the drag force, or too large deviations from our assumptions, for example only applying the drag force to the companion and not the core of the RSG. The core has a mass of $2.8 \, \mathrm{M}_\odot$ which is lower than the mass of the BH companion, meaning that the center of mass, around which the binary orbits, is no longer close to the core the of RSG but closer to the companion. 

The envelope ejection only starts later in the evolution, similar to the simulation with the NS mass companion. At around $t=1700\,\mathrm{d}$, we find a similar amount of ejected envelope and similar ejection rates compared to the 3D simulations. However, because the 1D simulations need to be stopped shortly after, we cannot compare any subsequent mass ejection.

Both the simulation with the NS mass companion and the BH mass companion reach only partial envelope ejection with $50\%$ and $25\%$ of the envelope ejected respectively. Therefore, we cannot make any statements on the final outcome of the CE evolution, that is complete envelope ejection or the merger of the companion and the core of the RSG.

\subsection{Comparison between AGB and RSG CE simulations}\label{sec:res_comp}

\begin{figure}
    \centering
    \includegraphics[width=12cm]{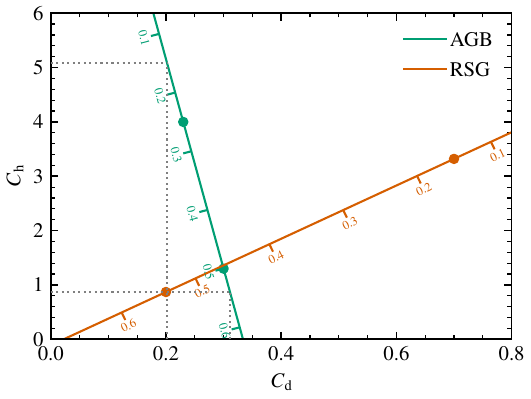}
    \bigskip
    \begin{minipage}{12cm}
    \centering
        \caption{Extrapolations for $C_\mathrm{d}$ and $C_\mathrm{h} $ for the best-fitting simulations with the AGB and the RSG donor. Ticks along the lines indicate the corresponding mass ratio $q$. The expected values for $C_\mathrm{d}$ and $C_\mathrm{h}$ for the RSG donor based on the AGB extrapolation are indicated by the dotted lines.}
        \label{fig:CdCh_comp}
    \end{minipage}
\end{figure}

We find well-fitting simulations for two mass ratios $q$ for both the CE simulations of the AGB donor and the RSG donor. Based on these, we test how much the obtained values for the drag-force parameter $C_\mathrm{d}$ and the heating parameter $C_\mathrm{h}$ vary. We perform linear interpolation/extrapolation of $C_\mathrm{d}$ and $C_\mathrm{h}$, using the mass ratio $q$ as the interpolation/extrapolation variable (Fig.~\ref{fig:CdCh_comp}). Extrapolation from the AGB simulation would suggest $C_\mathrm{d} = 0.20 $ and $C_\mathrm{h} = 5.08$ for the RSG with the NS mass companion. However, we are only able to reproduce the 3D simulations when using a much larger drag-force parameter. For the BH mass companion, the AGB simulations suggest $C_\mathrm{d} = 0.31$ and $C_\mathrm{h} = 0.87$, which are much closer to the values we obtained from fitting to the 3D simulations. We conclude, that a calibration of $C_\mathrm{d}$ and $C_\mathrm{h}$ on the 3D simulations, if possible, depends on the envelope structure of the donor star. 

The qualitative behavior of the spiral-in curve and the amount of ejected envelope is similar between the AGB and RSG donor simulations. We consistently find that envelope ejection starts delayed compared to the 3D simulation. This is caused by a different envelope ejection mechanism in the 1D simulations \citep{Bronner2024}. The CE simulation of the RSG with the BH mass companion is the only case where we find a larger post-plunge-in separation in the 1D simulation compared to the 3D simulation. Drawing any conclusion from this is too early because we only have five 1D CE simulations available. Further 1D simulations in the future will help to understand this better.

From the 1D simulation with the AGB donor, we concluded that the recombination energy released from hydrogen plays an important role in expanding the envelope \citep{Bronner2024}. In comparison, we find that for the RGS donor, the recombination energy released from He~{\sc iii} might be the more dominant energy source. In the RSG case, the recombination zones of hydrogen and He~{\sc ii} are located close together, while we found a clear separation for the recombination zones in the AGB case. The difference in the ionization or recombination structure of the AGB and RSG seems to be the most prominent difference between the CE evolution of low-mass stars and massive stars, although it is not yet clear whether these differences are important for the CE evolution.

\section{Conclusions and outlook}\label{sec:conc}
We extend the work presented in \citet{Bronner2024} to a $9.2\, \mathrm{M}_\odot$ RSG donor with an NS mass and BH mass companion. The 1D CE simulations are able to reproduce the results of \citet{Vetter2024}. This shows that our 1D CE method can be used for both CE phases of low-mass stars and massive stars. However, as the mass of the companion exceeds a critical value, the 1D method can no longer be used, as the assumptions break down. The critical mass is most likely related to the core mass of the giant star and is of similar magnitude. Exceeding this mass breaks down the assumption of the 1D approach. We find that the ionization structure in the RSG envelope evolves differently from the AGB envelope. The recombination energy of hydrogen seems to play less of a role compared to the AGB CE simulation. A comparison between the values for $C_\mathrm{d}$ and $C_\mathrm{h}$ for the best-fitting simulation already indicates, that a global calibration, if possible at all, will depend on the structure of the giant star.

In future work, we plan to modify the numerical setup to prolong the 1D simulation such that we are able to reach either full envelope ejection or a merger of the companion with the core of the giant star. Additionally, we plan to repeat similar simulations with many more donor stars. Using the best-fitting models when comparing to 3D simulations, we might be able to find a global calibration for $C_\mathrm{d}$ and $C_\mathrm{h}$. If such a calibration can be found, the model can be used for predictions for post-CE properties, such as the post-CE orbital separation, or to decide whether a CE event leads to a successful ejection of the envelope or a merger. In principle, a map between $(C_\mathrm{d},C_\mathrm{h})$ of our 1D method and $(\alpha_\mathrm{CE},\lambda)$ of the CE energy formalism could be possible. However, a dense sampling of 1D CE simulations would be required to construct such a map, as this cannot be done from first principles. Nonetheless, if such a map could be constructed, the classical energy formalism could be used to determine the outcome of the CE phase. This might then be used to improve the CE physics in population synthesis calculations or 1D binary evolution calculations.


% ...

\begin{acknowledgments}
V.A.B.~thanks Marco Vetter for providing detailed results of \citet{Moreno2022} and \citet{Vetter2024}, and fruitful discussions regarding CE evolution.
The authors acknowledge support from the Klaus Tschira Foundation.
This work has received funding from the European Research Council (ERC) under the European Union's Horizon 2020 research and innovation programme (Grant agreement No.\ 945806). This work is supported by the Deutsche Forschungsgemeinschaft (DFG, German Research Foundation) under Germany's Excellence Strategy EXC 2181/1-390900948 (the Heidelberg STRUCTURES Excellence Cluster).
F.K.R.\ acknowledges funding by the European Union (ERC, ExCEED, project number 101096243). Views and opinions expressed are however those of the authors only and do not necessarily reflect those of the European Union or the European Research Council Executive Agency. Neither the European Union nor the granting authority can be held responsible for them.
\end{acknowledgments}

\begin{furtherinformation}

\begin{orcids}

\orcid{0000-0002-7624-2933}{Vincent~A.}{Bronner}
\orcid{0000-0002-5965-1022}{Fabian~R.~N.}{Schneider}
\orcid{0000-0002-8338-9677}{Philipp}{Podsiadlowski}
\orcid{0000-0002-4460-0097}{Friedrich~K.}{R\"opke}
\end{orcids}

\begin{authorcontributions}
    \textit{Conceptualization}: F.N.R.S., Ph.P.;
    \textit{Data curation}: V.A.B.;
    \textit{Formal analysis}: V.A.B.;
    \textit{Funding acquisition}: F.N.R.S.;
    \textit{Investigation}: V.A.B.;
    \textit{Methodology}: V.A.B., F.R.N.S, Ph.P., F.K.R.;
    \textit{Project administration}: F.N.R.S.;
    \textit{Resources}: F.N.R.S.;
    \textit{Software}: V.A.B.;
    \textit{Supervision}: F.N.R.S.;
    \textit{Validation}: V.A.B.;
    \textit{Visualization}: V.A.B.;
    \textit{Writing -- original draft}: V.A.B.;
    \textit{Writing -- review \& editing}: V.A.B., F.N.R.S., Ph.P., F.K.R.
\end{authorcontributions}

\begin{conflictsofinterest}
    The authors declare no conflict of interest
\end{conflictsofinterest}

\end{furtherinformation}



%

\bibliographystyle{bullsrsl-en}

\bibliography{VA_BRONNER}

\end{document}